\documentclass[11pt,a4paper]{article}
\usepackage{jheppub}
\usepackage[T1]{fontenc}
\usepackage{amsmath}
\usepackage{amssymb}
\usepackage{esint}
\usepackage{amsfonts}
\usepackage{graphicx,color}
\usepackage{slashed}
\usepackage{young}
\usepackage[vcentermath]{youngtab}
\usepackage[utf8]{inputenc}
\usepackage{tensor}
\usepackage{etoolbox}
\usepackage{bbm}
\usepackage{soul}
\usepackage{ulem}

\makeatletter

\newcommand{\be}{\begin{equation}}
\newcommand{\ee}{\end{equation}}
\newcommand{\bea}{\begin{eqnarray}}
\newcommand{\eea}{\end{eqnarray}}

\renewcommand{\tilde}{\widetilde}
\renewcommand{\hat}{\widehat}
\newtheorem{prop}{Proposition}[section]

\newtheorem{theorem}[prop]{Theorem}

\renewcommand{\d}{\partial}

\def\cL{\mathcal{L}}

\newcommand*\xbar[1]{%
  \hbox{%
    \vbox{%
      \hrule height 0.5pt 
      \kern0.3ex
      \hbox{%
        \kern-0.0em
        \ensuremath{#1}%
        \kern-0.0em
      }%
    }%
  }%
} 

\pdfpageheight\paperheight
\pdfpagewidth\paperwidth

\pdfoutput=1 

    \patchcmd{\maketitle}{\@fpheader}{}{}{}

\title{Asymptotic symmetries of electromagnetism at spatial infinity}
\author[a,b]{Marc Henneaux,}
\author[c]{and C\'edricTroessaert}
\affiliation[a]{Universit\'e Libre de Bruxelles and International Solvay Institutes, ULB-Campus Plaine CP231, B-1050 Brussels, Belgium}
\affiliation[b]{Coll\`ege de France, 11 place Marcelin Berthelot, 75005 Paris, France}
\affiliation[c]{Max-Planck-Institut f\"{u}r Gravitationsphysik (Albert-Einstein-Institut),
Am M\"{u}hlenberg 1, \\ DE-14476 Potsdam, Germany}

\abstract
{We analyse the asymptotic symmetries of Maxwell theory at spatial infinity through the Hamiltonian formalism. Precise, consistent boundary conditions are explicitly given and shown to be invariant under asymptotic angle-dependent $u(1)$-gauge transformations.  These symmetries generically have  non-vanishing charges.  The algebra of the canonical generators of this infinite-dimensional symmetry with the Poincar\'e charges is computed.  The treatment  requires the addition of surface degrees of freedom at infinity and a modification of the standard symplectic form by surface terms.  We extend the general formulation of well-defined generators and Hamiltonian vector fields to encompass such boundary modifications of the symplectic structure.  Our study covers magnetic monopoles.}

\makeatother

\begin{document}
\maketitle \flushbottom

\section{Introduction}
\setcounter{equation}{0}

Recent work on the asymptotic structure of electromagnetism in Minkowski spacetime has revealed the remarkable presence of an infinite-dimensional symmetry at infinity, which enables one to view the soft photon theorems in quantum electrodynamics as the associated Ward identities  \cite{Balachandran:2013wsa,Strominger:2013lka,Barnich:2013sxa,Bern:2014vva,He:2014cra,Lysov:2014csa,Kapec:2014zla,Kapec:2015ena,Campiglia:2016hvg,Conde:2016csj,Campiglia:2017mua} (for a review of these fascinating developments, see \cite{Strominger:2017zoo}).  

Most of these studies were carried out at null infinity. The purpose of this paper is to shed light on these questions by performing the analysis at spatial infinity.  A description of the symmetry and its associated charges on standard spacelike hyperplanes adapted to inertial Lorentz observers, who have access to the information available at null infinity since their hyperplanes of simultaneity are Cauchy surfaces, must indeed be possible -- and is in fact so, as we shall show. Our work also resolves the tension between the absence of effective infinite symmetry at spatial infinity found in some earlier work and the recent developments mentioned above.

In order to perform the analysis at spacelike infinity, one must provide there precise boundary conditions on the dynamical variables. These conditions should be consistent, i.e., should fulfill the following criteria:
\begin{itemize}
\item They should make the action, and in particular the symplectic form, finite.
\item They should be such that all Poincar\'e transformations are symmetries of the theory.  That is, the Poincar\'e transformations, including the dynamical ones (time translations and Lorentz boosts), should be canonical transformations that leave the boundary conditions invariant and have well-defined (finite) canonical generators.
\item They should include the ``physically'' relevant solutions (in particular the Coulomb potential for a charge at rest).
\end{itemize}

Boundary conditions fulfilling these criteria were given in
\cite{Henneaux:1999ct}.  These boundary conditions involve parity conditions
inspired by \cite{Regge:1974zd} and are invariant under an
infinite-dimensional set of angle dependent $u(1)$ transformations  fulfilling
also some definite parity conditions.  However,  the corresponding charges
turn out to vanish for all configurations obeying the boundary conditions,
indicating that they are ``proper gauge transformations''
\cite{Benguria:1976in}  not changing the physical state of the system.  This
shows that the analysis of \cite{Henneaux:1999ct} must be completed   and that
the problem of finding consistent boundary conditions allowing for a non
trivial action of the asymptotic infinite-dimensional symmetries requires delicate investigation.

A similar problem arises for pure gravity, where the BMS group \cite{Bondi:1962px,Sachs:1962wk,Sachs:1962zza,Newman:1962cia,Penrose:1962ij,Penrose:1965am,Madler:2016xju,Alessio:2017lps} investigated at null infinity is absent with the parity conditions of \cite{Regge:1974zd}.  The difficulty was solved in 
\cite{Henneaux:2018cst} where new boundary conditions involving a crucial  ``twist'' of the parity conditions for the angular components of the fields were proposed and shown to yield a non trivial action of the BMS group. 

We show in this paper that a similar twist with respect to the parity conditions of \cite{Henneaux:1999ct} is necessary in the electromagnetic case.  We also prove that it is in fact sufficient that this twist be an improper gauge transformation for the emergence of the non-trivial infinite-dimensional symmetry.  What is crucial is to allow {\em non zero} twisted parity components of the angular components of the potential.  The virtue of taking these twisted components to be simply an improper gauge transformation and not arbitrary functions with twisted parity,  is that one can then easily cover magnetic sources.  Furthermore, this restricted form of the twist eliminates singular behaviour of some of the fields as one tends to null infinity.

We also investigate Lorentz invariance, known to be a subtle issue due to the long-range features of the electromagnetic field \cite{Frohlich:1979uu,Frohlich:1978bf}, already at the classical level (see e.g. \cite{Herdegen:1995nf} and \cite{Balachandran:2013wsa} and references therein). 
The most unexpected feature coming out of our analysis is the manner in which  the Lorentz boosts are canonically realized.  Indeed, these fail to be canonical transformations, unless one adds a surface degree of freedom at infinity.  This degree of freedom contributes to the symplectic structure,  which acquires therefore an extra surface term.   Boundary contributions to the symplectic form  were discussed earlier in \cite{Ashtekar:2000hw} in the different context of isolated horizons.  Here, our extra surface term in the symplectic structure reproduces the surface contribution of \cite{Campiglia:2017mua} when one fixes the gauge by relating the temporal component $A_0$ of the vector potential to the new surface degree of freedom. However, we prefer not to fix the gauge in our general derivation and leave $A_0$ unrelated to the surface degree of freedom.  Once the new degree of freedom and the surface contribution have been included, the Lorentz group is canonically realized.  

The idea of introducing surface degrees of freedom  was pursued previously in  \cite{Gervais:1976ec,Wadia:1976fa,Wadia:1977qr,Gervais:1980bz} to describe the infrared sector of gauge theories, and more recently in \cite{Dvali:2015onv} where they appear as holographic St\"uckelberg fields, but we have not explored
the connection (if any) between the variables introduced there and the variables introduced here.

Our paper is organized as follows.  The next section, Section \ref{sec:General}, recalls some background information on electromagnetism in Minkowski space. The explicit action of the Poincar\'e group on the leading orders of the asymptotic fields is given and the asymptotic form of some key solutions (Li\'enard-Wichiert potentials, magnetic monopoles) is given.    We then show in Section \ref{subsec:ParityC} that the finiteness of the bulk symplectic form imposes extra conditions on the asymptotic fields.  Combined with the known behaviour of the key solutions, this leads to a definite set of asymptotic conditions.  The asymptotic behaviour involves a twist of the parity conditions for the angular part of the fields, which is crucial for the emergence of a non trivial infinite-dimensional asymptotic symmetry. Section \ref{sec:EMaction} constitutes the core of our paper.  It deals with the zero magnetic charge case, which illustrates all the main points. We first exhibit the need to modify the symplectic form by boundary contributions which involve new surface degrees of freedom in order for the boosts to be canonical transformations. We then provide the complete action, with these new degrees of freedom included.  The global symmetries are worked out and shown to involve an angle-dependent $u(1)$ algebra. 
The treatment  is extended in Section \ref{sec:Magnetic} to cover magnetic sources. The extension is direct. Indeed, all the new conceptual points are already present in the zero magnetic charge case of Section \ref{sec:EMaction} since they are related to the twist in the parity conditions. Section \ref{sec:Conclusions} is devoted to concluding comments and open questions. In particular, the way to cover Taub-NUT and magnetic charges in the Einstein-Maxwell system is indicated. Finally, Appendix \ref{sec:PoissonStructure} extends the general formulation of well-defined generators and Hamiltonian vector fields to cover boundary modifications of the symplectic structure, while Appendices \ref{sec:EMhyperbolic} and \ref{app:u1transfos} provide the explicit link between the asymptotic symmetry algebras at null infinity and spatial infinity, and demonstrate their equality.

\section{Some background}

\label{sec:General}

\subsection{Action and gauge symmetries}
\label{subsec:Starting}

We start with free electromagnetism in Minkowski space, in standard Minkowskian coordinates.  The dynamical variables to be varied in the action are the spatial components $A_i$ of the vector potential,  their conjugate momenta $\pi^i$ (equal to the electric field) and the temporal component $A_0 \equiv A_t$ of the vector potential which plays the role of Lagrange multiplier  for the constraint
\be
\mathcal{G} = - \partial_i \pi^i \approx 0
\ee
(Gauss' law).   We
use the symbol $\approx$ to denote equality on the constraint's surface. The action is 
\begin{equation}
\label{eq:StartingPoint}
	S_H[A_i, \pi^i, A_0] = \int dt \left\{ \int d^3x \, 
		\pi^i \d_t
		A_i 
		 	- \int d^3x \left(   
				\frac{1}{2} \pi^i \pi_i + \frac{1}{4} F^{ij} F_{ij}  + A_t  \mathcal{G}\right) + F_\infty\right\}
\end{equation}
where $F_\infty$ is a surface term at spatial infinity ($r \rightarrow \infty$),
which depends on the boundary conditions and which will be discussed below.

One could couple charges (massive or massless) to the electromagnetic field
but we can assume that their fields decay sufficiently fast at spatial
infinity that they do not directly contribute to the surface integrals.  They
do, however, indirectly contribute by changing the fluxes.  For instance,  if
Gauss' law with zero right hand side holds everywhere in space ($\simeq
\mathbb{R}^3$), the flux of the electric field at spatial infinity vanishes.  The presence of charges replaces $\mathcal{G} \equiv - \partial_i \pi^i \approx 0 $ by $\mathcal{G} \equiv - \partial_i \pi^i  + j^0 \approx 0 $ where $j^0$ is the charge density, and modifies the flux of the electric field at infinity, which is no longer zero.  Assuming that the electric flux at infinity is non zero is the way we shall take the charges into account.

The electromagnetic field and its conjugate are usually taken to possess the following decay at  spatial infinity,
\begin{equation}
	A_i = \frac{1}{r} \xbar A_i + \frac{1}{r^2} A^{(1)}_i +
	o(r^{-2}),\quad
	\pi^i = \frac{1}{r^2} \xbar \pi^i + \frac{1}{r^3} \pi^{(1)i} +
	o(r^{-3}) \label{GenFallOff}
\end{equation}
where the coefficients of the various powers of $r^{-1}$ are arbitrary functions on the $2$-sphere, i.e., of the angles $x^A$ used to parametrize it\footnote{We shall assume ``uniform smoothness'' \cite{Sachs:1962zza} whenever needed, i.e., $\partial_r o(r^{-k}) = o(r^{-k-1})$, $\partial_A o(r^{-k}) = o(r^{-k})$.}. We shall find it necessary to strengthen  (\ref{GenFallOff}).  Extra constraints are indeed needed since the conditions (\ref{GenFallOff}) by themselves do not guarantee finiteness of the symplectic structure.  So, the fall-off conditions (\ref{GenFallOff}) are not the final boundary conditions. The form of the extra constraints  will be explicitly described in Section \ref{subsec:ParityC} below.

The boundary conditions (\ref{GenFallOff}) are invariant under gauge transformations generated by 
the first-class constraint-generator $\mathcal{G}$ :
\begin{equation}
	\delta_{\epsilon} A_i =  \d_i \epsilon, \quad
	\delta_{\epsilon} \pi^i = 0,  \label{eq:GaugeTrans0}
\end{equation}
provided the gauge parameter $\epsilon$ has the asymptotic behaviour
\begin{equation}
	\epsilon = \xbar \epsilon(x^A) + \frac{1}{r} \epsilon^{(1)}(x^A) +
	o(r^{-1}). \label{eq:GaugePar0}
\end{equation}
Further conditions limiting the functional class to which $\epsilon$ belongs will of course appear when strengthening the boundary conditions (\ref{GenFallOff}). The generator of (\ref{eq:GaugeTrans0}) is explicitly
\begin{equation}
	G[\epsilon] = \int d^3x \, \epsilon \, \mathcal{G} + \oint d^2S_i \, \xbar \epsilon \, \xbar \pi^i  \approx
	\oint d^2S_i \, \xbar \epsilon \, \xbar \pi^i   \label{eq:ChargeG0}
\end{equation}
where the surface term is determined by the methods of \cite{Regge:1974zd}.  The electric charge is associated
with the improper gauge transformation generated by  $\epsilon=  1$.  We recall that an ``improper'' gauge transformation \cite{Benguria:1976in} changes the physical state of the system and has non-vanishing charge.  It should not be quotiented out.  

To complete the description of the asymptotic behaviour, we need to specify the fall-off of the Lagrange multiplier $A_t$.  Since   $A_t$ parametrizes the gauge transformation performed in the course of the evolution, we take for $A_t$ the same fall-off as for the gauge parameter $\epsilon$, 
\begin{equation}
	A_t = \xbar A_t(x^A) + \frac{1}{r} A_t^{(1)}(x^A) +
	o(r^{-1}).
\end{equation}
If $\xbar A_t(x^A) = C \not=0$,  the time evolution involves a non-trivial
improper gauge transformation.  The term $\xbar A_t(x^A)$ will be subject to the same extra conditions as the gauge parameter $\epsilon$ when strengthening the boundary conditions.

\subsection{Poincar\'e transformations}

We now turn to Poincar\'e transformations.   A general deformation of a spacelike  hyperplane can be decomposed into normal and tangential components, denoted by $\xi$ and $\xi^i$, respectively. A general Poincar\'e transformation corresponds to the deformation
\begin{eqnarray}
 \xi &=& b_i x^i + a^\perp  \label{eq:Poinc1}\\
 \xi^i &=&{ b^i}_j x^j + a^i  \label{eq:Poinc2}
\end{eqnarray}
where $b_i$, $b_{ij} = -b_{ji}$, $a^\perp$ and $a^i$ are arbitrary constants. The constants $b_i$ parametrize the Lorentz
boosts, whereas the antisymmetric constants $b_{ij} = -b_{ji}$
parametrize the spatial rotations.  The constants $a^\perp$ and $a^i$ are standard translations. 

Under such a deformation, the fields transform as
\begin{eqnarray}
&&\delta A_i = \xi \pi_i  + \xi^j F_{ji}  + \partial_i \zeta \label{eq:GTA}\\
&& \delta \pi^i = \partial_m \left( F^{mi} \xi \right) + \partial_m \left(\xi^m \pi^i \right) - (\partial_m \xi^i )\pi^m - \xi^i \partial_m \pi^m .
\end{eqnarray}
The transformation of the fields is really defined up to a gauge transformation.  This is the reason why we have included the term $\partial_i \zeta$ in the transformation of $A_i$.  A definite choice  of accompanying gauge transformation will be made below to get simple expressions for the algebra.
It is clear that the fall-off (\ref{GenFallOff}) is preserved under these  transformations provided $\zeta$ behaves as in \eqref{eq:GaugePar0}.

For later purposes, we rewrite the boundary conditions in spherical coordinates.  One gets, recalling that the momenta carry a unit density weight:
\begin{gather}
	\label{eq:hamilasymptI0}
	A_r = \frac{1}{r} \xbar A_r + \frac{1}{r^2} A^{(1)}_r +
	o(r^{-2}),\quad
	\pi^r = \xbar \pi^r + \frac{1}{r} \pi^{(1)r} +
	o(r^{-1}),\\
	A_A = \xbar A_A + \frac{1}{r} A^{(1)}_A +
	o(r^{-1}),\quad
	\pi^A = \frac{1}{r}\xbar \pi^A + \frac{1}{r^2} \pi^{(1)A} +
	o(r^{-2}), \label{eq:hamilasymptI022}
\end{gather}
while the form of $A_t$ remains unchanged as it is a spatial scalar.  The coordinates $x^A$ are coordinates on the two-sphere. We will also need the Poincar\'e transformations of the leading orders, which  are invariant under translations and transform only under boosts and spatial rotations.  Recalling that the above transformations  read
$ \delta A_i = \frac{ \xi \pi_i }{\sqrt{g}} + \xi^j F_{ji} + \partial_i \zeta$ and 
$ \delta \pi^i = \partial_m \left( \sqrt{g} F^{mi} \xi \right) + \partial_m \left(\xi^m \pi^i \right) - (\partial_m \xi^i )\pi^m - \xi^i \partial_m \pi^m$ in general curvilinear coordinates, where $g_{ij}$  is the flat metric in those coordinates,  one finds explicitly
\begin{gather}
	\delta_{b,Y} \xbar A_r = \frac{b}{\sqrt{\xbar \gamma}} \xbar \pi^r +
	Y^A \d_A \xbar A_r,  \label{eq:TransLead2}\\
	\delta_{b,Y} \xbar A_A = \frac{b}{\sqrt{\xbar\gamma}} \xbar
	\gamma_{AB}\xbar \pi^B + Y^B \left(\d_B \xbar A_A -   \d_A \xbar
	A_B\right) + \partial_A \xbar \zeta, \label{eq:TransLead2bis}\\
	\delta_{b,Y} \xbar \pi^r = \sqrt {\xbar \gamma}\, \xbar D^A(b \d_A \xbar A_r) 
	+ \d_A(Y^A \xbar \pi^r), \label{eq:TransLead3}\\
	\delta_{b,Y} \xbar \pi^A = \sqrt {\xbar \gamma}\,\xbar D_B \Big(b\, \xbar \gamma^{BC}
	\xbar \gamma^{AD}(\d_C \xbar A_D - \d_D \xbar A_C)\Big) \nonumber \\  \hspace{1.5cm} + \d_B(Y^B \xbar \pi^A) - \d_B Y^A \xbar
	\pi^B - Y^A \partial_B \xbar \pi^B \label{eq:TransLead4}
\end{gather}
where we have set, in terms of the unit metric 
$\xbar \gamma_{AB}$ on the sphere, 
\be
g_{AB} = r^2\xbar\gamma_{AB}
\ee
and
\begin{gather}
	\xi = rb + T,\quad \xi^r = W, \quad \xi^A = Y^A + \frac 1 r \xbar D^A
	W,  \label{eq:FormOfW}\\
	\xbar D_A \xbar D_B W + \xbar \gamma_{AB} W = 0, \quad
	\xbar D_A \xbar D_B b + \xbar \gamma_{AB} b = 0, \quad \cL_Y \xbar
	\gamma_{AB} = 0, \quad \d_A T = 0.
\end{gather}
The quantities $b$, $Y^A$, $T$ and $W$ are functions on the sphere.  The first
two, $b$ and $Y^A$, describe the homogeneous Lorentz transformations, while
$T$ and $W$, which do not appear in the transformation laws
(\ref{eq:TransLead2})-(\ref{eq:TransLead4}) of the leading orders, describe the translations.  Explicitly,  
\be
b = b_1 \sin \theta \cos \varphi + b_2 \sin \theta \sin \varphi + b_3 \cos \theta,
\ee
\be
\hspace{-.5cm} Y = m^1 \left(-\sin \varphi \frac{\partial}{\partial \theta} - \frac{ \cos \theta}{\sin \theta} \cos \varphi \frac{\partial}{\partial \varphi}\right)+ m^2 \left(\cos \varphi \frac{\partial}{\partial \theta} - \frac{ \cos \theta}{\sin \theta} \sin \varphi \frac{\partial}{\partial \varphi}\right) + m^{3} \frac{\partial}{\partial \varphi}
\ee
and
\be
W = a^1 \sin \theta \cos \varphi + a^2 \sin \theta \sin \varphi + a^3 \cos \theta.
\ee
Finally,  $\xbar D_A$ is the covariant derivative associated with 
$\xbar \gamma_{AB}$  and $\xbar D^A = \xbar  \gamma^{AB} \xbar D_B$.

The transformation laws (\ref{eq:TransLead2})-(\ref{eq:TransLead4}) of the leading orders possess interesting features:  
\begin{itemize}
\item They do not mix radial and angular components.  The radial variables transform among themselves and so do also the angular ones.  This implies that one can treat independently the radial and angular components in the boundary conditions.
\item Since $b$ and $Y^A$ are odd under the parity transformation $x^i
	\rightarrow - x^i$, which we formally write as $x^A \rightarrow - x^A$
	in terms of the coordinates on the sphere\footnote{Note that in terms
	of standard spherical coordinates, the antipodal map is actually
$\theta \rightarrow \pi - \theta$ and $\varphi \rightarrow \varphi + \pi$ (and
$r \rightarrow r$).  This implies $d \theta \rightarrow - d \theta$ and $d
\varphi \rightarrow d \varphi$.  Therefore, the condition that $A_A$ is even
(for example), i.e., $A_A (-x^B) =  A_A(x^B)$, which really means that the
one-form $A = A_A dx^A$ is odd (i.e., $\Phi^* A = -A$, where $\Phi^*$ denotes
the pullback by the antipodal map), is equivalent to the statement that
$A_\theta$ is even and $A_\varphi$ is odd. Similar considerations apply to the
odd case $A_A (-x^B) =  -A_A(x^B)$. \label{foot:Parity}}, one can consistently
impose parity conditions on the leading orders of the canonical variables.
That is, one can require the components of the potential  to have a definite
parity and their conjugate momenta to have the opposite one.  This is
preserved by the Lorentz transformations (if one chooses the gauge
transformation $\zeta$ appropriately).  Furthermore, one can consider different parity conditions for the radial and angular components since they transform independently.
\end{itemize}

In polar coordinates, the conserved charge (\ref{eq:ChargeG0}) takes the form
\begin{equation}
	G[\epsilon]  \approx
	\oint d^2x \, \xbar \epsilon \, \xbar \pi^r  .\label{eq:ChargeG1}
\end{equation}
 It follows from our boundary conditions that $\xbar A_r$ is gauge invariant since $\partial_r \epsilon = O(r^{-2})$.  The gauge invariance of $\xbar A_r$ can be phrased in familiar terms by observing that the radial integral (along any fixed ray) $\int_R^{2R} dr A_r$ transforms as $\epsilon (2R) - \epsilon (R) = O(\frac1R)$ and is thus gauge invariant in the limit $R \rightarrow \infty$. This integral is equal to $ \xbar A_r \ln2$ in that limit.  The symmetry generated by $\xbar A_r$ will be clarified below.

\subsection{Li\'enard-Wiechert solution}

To motivate the boundary conditions below, we now consider various solutions.  We start with the Coulomb potential.

The Coulomb solution has $\xbar A_r = 0$, $\xbar \pi^r = \sin \theta $, $\xbar A_A = 0$ and $\xbar \pi^A = 0$.  By boosting it, one generates the Li\'enard-Wiechert solution.   Since $\xbar \pi^r$ is even, the Li\'enard-Wiechert solution is characterized by the following parities of the radial components, 
\be
\xbar A_r(-x^A)  = - \xbar A_r(x^A)  , \qquad \xbar \pi^r(-x^A)  =  \xbar \pi^r(x^A).
\ee
The statement that $\xbar A_r$ is odd is gauge invariant because $\xbar A_r$ is gauge invariant.

As to the angular components, both $\xbar \pi^A$ and $\xbar F_{AB}$ remain zero,
\be \xbar \pi^A = 0, \qquad \xbar F_{AB}= 0. 
\ee
 This implies that $\xbar A_A$ reduces to a gauge transformation $\partial_A \xbar\Phi$,
\be
\xbar A_A = \partial_A \xbar\Phi 
\ee
for some function $\xbar\Phi$ of the angles.  The value of $\xbar\Phi$ depends on the choice of the accompanying gauge transformation $\zeta$ in (\ref{eq:TransLead2})-(\ref{eq:TransLead4}).  The choice $\zeta = 0$ amounts to transform $A_\mu$ as $\xi^\rho F_{\rho \mu}$ under Poincar\'e transformations.  One other possible choice is to take $\zeta$ so that the transformation of the potential $A_\mu$ is its Lie derivatives $\mathcal L_\xi A_\mu$ rather than $\xi^\rho F_{\rho \mu}$. The two choices differ by the gauge transformation $\partial_{\mu} (\xi^\rho A_\rho)$.  The gauge parameter $\xi^\rho A_\rho$  is of order $O(1)$ for rotations and boosts.

\subsection{Magnetic monopoles}
For magnetic monopoles, the situation is somewhat reversed. The asymptotic Coulomb fields $\xbar A_r$ and $\xbar \pi^r$ vanish and remain zero under Poincar\'e transformations.  The field of a monopole is purely angular and given by
\be
\pi^A = 0 , \qquad F_{\theta \varphi} = \sin \theta
\ee
when the monopole is at rest. It coincides with its leading order. The 2-form $\frac12  \xbar F_{AB} dx^A dx^B$ is odd, something that we write as $\xbar F_{AB}(-x^C) = - \xbar F_{AB}(x^C)$.  By Poincar\'e transforming the monopole field, one generates $\xbar \pi^A$ and $\xbar F_{AB}$ that remain odd. The potential is not globally defined but on the sphere minus the two (antipodal) poles, one can take it to be even up to a gauge transformation, e.g., for the monopole at rest
\begin{equation}
	\xbar A_\theta = \partial_\theta \xbar\Phi, \quad \xbar A_\varphi = -
	\cos \theta + \partial_\varphi \xbar\Phi
\end{equation}
(see footnote {\footnotesize{\ref{foot:Parity}}} for conventions on parity
terminology).  The (not globally defined) gauge transformation that brings the
monopole potential to the familiar form $\xbar A_\varphi = (1- \cos \theta) $ regular at the North pole has gauge parameter $\epsilon =\varphi$ and is such that the one-form $d \epsilon = d \varphi$ is odd (opposite parity to that of $ - \cos \theta d \varphi$).

\section{How to make the symplectic form finite}
\label{subsec:ParityC}

Without further constraints, the symplectic potential derived from the bulk piece of (\ref{eq:StartingPoint})  is logarithmically divergent, since its dominant part is
\be
\int \frac{dr}{r} \int d^2x \left(\xbar \pi^r \dot{ \xbar A_r} + \xbar \pi^A
\dot{ \xbar A_A} \right).
\ee 
To make it convergent, one must impose extra conditions on the leading components of the dynamical variables so that the integral on the $2$-sphere
\be
\int d^2x \left(\xbar \pi^r \dot{ \xbar A_r} + \xbar \pi^A \dot{ \xbar A_A} \right)
\ee
vanishes.  

We treat separately the radial and angular components.  

\subsection{Condition on the radial components}

To make the symplectic form finite for the radial components, we impose a
 parity condition, as for gravity \cite{Regge:1974zd}.    In view of our above discussion, we request 
\be
\xbar A_r(-x^A)  = - \xbar A_r(x^A)  , \qquad \xbar \pi^r(-x^A)  =  \xbar \pi^r(x^A) \label{eq:parityRadial}
\ee
This agrees with \cite{Henneaux:1999ct}.  As we have seen, this boundary condition contains the Coulomb field viewed in a moving frame.  The antipodal symmetry of $\xbar \pi^r$ is what remains of the spherical symmetry of the static field after an arbitrary Poincar\'e transformation has been performed.  As we have also stressed, (\ref{eq:parityRadial}) is gauge invariant.  The parity condition (\ref{eq:parityRadial}) will be imposed throughout the subsequent discussion.

Since $\xbar\pi^r$ is even, the charges (\ref{eq:ChargeG1}) reduce to
\be G[\epsilon]  \approx
	\oint d^2x \, \xbar \epsilon_{\hbox{\tiny even}} \, \xbar \pi^r . \label{eq:ChargeG2}
\ee
The odd part $\epsilon_{\hbox{\tiny odd}}$ of the gauge parameter $\epsilon$ gives a zero contribution to the charges $G[\epsilon]$.  It follows that the gauge transformations with $\epsilon_{\hbox{\tiny odd}}$ are proper gauge transformations that do not change the physical state of the system.  By contrast, the gauge transformations with $\epsilon_{\hbox{\tiny even}}$ are improper gauge transformations that do change the physical state of the system.  The charge (\ref{eq:ChargeG2}) is generically non zero.

These conclusions hold irrespectively of whatever extra conditions are imposed on the angular components of the fields to make $\int d^2x  \, \xbar \pi^A \dot{ \xbar A_A} $ vanish, conditions to which we now turn.

\subsection{Conditions on the angular components}

\subsubsection*{No magnetic charge}
In the absence of magnetic charges,  we impose the boundary conditions 
\be \xbar \pi^A = 0, \qquad \xbar A_A = \partial_A \xbar \Phi  \label{eq:BCNoMagn}\ee
for some function $\xbar \Phi$ of the angles.  As our previous discussion shows, these conditions are fullfilled by the Li\'enard-Wiechert potentials, and match therefore the setting adopted in most discussions of the behaviour of the fields at null infinity \cite{Strominger:2017zoo}. 

It might be tempting to set $\xbar \Phi$ equal to zero by a gauge transformation but this would be illegitimate if the needed gauge transformation is improper, which is the case when $\xbar \Phi$ is even.   For that reason, we keep $\xbar \Phi$ - or at least its even part.  The odd part of $\xbar \Phi$ defines instead a proper gauge transformation and can be set equal to zero if one so wishes, although it can be useful to keep it. 

Note that the relevant part of $\xbar A_A$ is odd since  $\xbar \Phi _{\hbox{\tiny improper}}$ is even.  We could therefore assume that $\xbar A_A$ is odd.  This means that $\xbar A_r dr$ and $\xbar A_A dx^A$ have opposite parities, which is the analog of the ``twisted parity conditions'' considered in \cite{Henneaux:2018cst} for gravity.  

\subsubsection*{Magnetic charges}

In the presence of magnetic poles, we keep the same conditions on the radial components of the fields so as to allow electric sources, and we impose on the angular components the parity conditions
\be \xbar A_A = \xbar A_A^{\hbox{\tiny even}} + \partial_A \xbar \Phi , \qquad \xbar \pi^A(-x^B)  =  -\xbar \pi^A(x^B) \label{eq:BCMagn1}
\ee
which are in agreement with the behaviour of the fields of magnetic sources.  Here
\be
\xbar A_A^{\hbox{\tiny even}}(-x^B)  =  \xbar A_A^{\hbox{\tiny even}}(x^B)  
\ee
Thus, the angular part of the potential is even up to a gauge transformation, while the angular part of the electric field is odd.  Again, one cannot drop the gauge component $\partial_A \xbar \Phi$ since its odd part defines an improper gauge transformation that changes the physical state.  Thus both the even and odd parts of $\xbar A_A$ are essential, and the odd part takes the form $\partial_A \xbar \Phi$.

To make the symplectic form finite, i.e., $\int d^2x  \, \xbar \pi^A \dot{ \xbar A_A} $ vanish, we impose also the Poincar\'e invariant condition
\be
\partial_A \xbar \pi^A = 0 \label{eq:BCMagn2}
\ee
Since $\d_r \pi^r = O(r^{-2})$, this condition is equivalent to demanding  that Gauss' law holds asymptotically, i.e.,  that the leading order $O(r^{-1})$ in $\d_i \pi^i$ be absent. This restriction has no physical
impact in the sense that it does not remove any solution, for which $\d_i \pi^i$ is strictly zero.  A similar condition was previously used for gravity in the study of the asymptotic symmetries of
asymptotically $AdS_3$ spacetimes in \cite{Brown:1986nw} and, more recently, in the
hamiltonian analysis of $BMS_4$ symmetries in \cite{Henneaux:2018cst}. 

It is clear that the action and the boundary conditions are invariant under
$u(1)$ transformations that  are even functions of the angles.  This
infinite-dimensional algebra acts non trivially.  The boundary conditions of
\cite{Henneaux:1999ct} assumed that $\xbar\Phi$ was odd so that $\xbar A_A$ was even. 
This choice corresponds to a uniform behaviour of the components of the vector potential and its conjugate in Cartesian coordinates, i.e.,
$\xbar A_i  =  \hbox{even} $, $ \xbar \pi^i  = \hbox{odd}$.
It is the analog for electromagnetism of the boundary conditions of \cite{Regge:1974zd}.
It freezes as we have seen the possibility of making improper gauge
transformations and explains why the non trivial asymptotic symmetry was found
to contain in that case only the global constant $u(1)$ transformations.  The
strict parity conditions  $\xbar A_A(-x^B) = \xbar A_A(x^B)$ of  \cite{Henneaux:1999ct} choose a definite point in the orbits of the angle-dependent $u(1)$ transformations.  To  be able to see the full orbits,  one must allow an odd part in the potential, as in (\ref{eq:BCMagn1}).

\section{Asymptotic analysis:  I. No magnetic charge}
\label{sec:EMaction}
\label{sec:EMhamilSym}

\subsection{Generalization of the boundary conditions}

We have exhibited so far an infinite-dimensional symmetry of electromagnetism characterized by an even function on the sphere.  The emergence of the other half of the symmetry, characterized by an odd function, is rather subtle.  The point is that the inclusion of an odd part in the vector potential, necessary as we have stressed to allow for the infinite symmetry, makes at the same time the action of the Lorentz transformations non canonical.  One way to cure this problem is to modify the symplectic structure by a boundary term involving a new surface degree of freedom.  The resulting action has a new symmetry involving the searched-for odd function on the sphere, which combines with the even function to yield the full symmetry displayed at null infinity.

The purpose of this section is to explain these somewhat unexpected features.  We start with the case of no magnetic charge with the boundary conditions (\ref{eq:BCNoMagn}) on the angular variables.  This  case illustrates already all the main features of the general construction.

It is in fact interesting to consider more general boundary conditions than
(\ref{eq:BCNoMagn}).  We shall assume in this section that the angular part of
the electric field $\xbar \pi^A$ does not vanish but is an even function on
the sphere that fulfills $\d_A \xbar \pi^A =0$, and that the angular part of
the vector potential is an odd function of the sphere modulo a gradient,
$\xbar A_A (-x^B) = - \xbar A_A(x^B) + \partial_A \xbar \Phi$.  The even part
of $\xbar \Phi$ can actually be absorbed in a redefinition of $A_A$ so that we
can assume $\xbar \Phi$ to be odd.  For that reason, one could drop $\xbar \Phi$ since it defines a proper gauge transformation, but we choose to keep it as it simplifies some formulas.  Thus, we take as boundary conditions in this section
\begin{eqnarray}
&&\xbar A_A(-x^B)  =  - \xbar A_A(x^B) + \partial_A \xbar \Phi, \qquad   \xbar
	\Phi(-x^B)  =  - \xbar \Phi(x^B), \label{eq:BcNew1} \\
&& \hspace{.5cm} \qquad \hbox{(Alternative -- and not the final! -- boundary conditions)} \nonumber \\
&& \xbar \pi^A(-x^B)  =  \xbar \pi^A(x^B) , \qquad \d_A \xbar \pi^A =0
\label{eq:BcNew2}
\end{eqnarray}
These boundary conditions make the symplectic form finite.

Because the angular components of the fields fulfill parity properties
different from those of the radial components, one calls (\ref{eq:BcNew1}) and
(\ref{eq:BcNew2})  ``twisted parity conditions". These boundary conditions
clearly contain (\ref{eq:BCNoMagn}) and hence,  accommodate the
Li\'enard-Wiechert potentials.  They also allow for improper gauge
transformations.  We shall carry the analysis of the no-magnetic-charge case
with these boundary conditions explaining along the way the simplifications
that occur if $\xbar A_A$ reduces to its  pure (improper) gauge transformation
piece and $\xbar\pi^A$ vanishes as in (\ref{eq:BCNoMagn}). 

There is an interest in carrying the analysis with the more general conditions (\ref{eq:BcNew1}) and (\ref{eq:BcNew2}) for various reasons.  First, it is always a good policy to devise boundary conditions as flexible as possible. Second, the analysis of (\ref{eq:BcNew1}) and (\ref{eq:BcNew2})  does not cost much more work than the analysis of the conditions (\ref{eq:BCNoMagn}).  Finally, the twisted parity conditions (\ref{eq:BcNew1}) and (\ref{eq:BcNew2}) are the analogs for electromagnetism of the twisted boundary conditions for gravity given in \cite{Henneaux:2018cst}.  Their analysis provides therefore useful insight into the limitations and properties of these boundary conditions.

\subsection{Lorentz boosts: need to add a boundary term to the symplectic structure}
\label{subsec:BoostsSymp}

For the Poincar\'e group to be a symmetry of the theory, it should leave the action invariant.  In particular, it should leave the symplectic structure invariant.  There is no difficulty with the bulk terms since the Poincar\'e transformations have canonical bulk generators and are thus formally canonical transformations.  The only subtlety comes from surface terms.   These are usually neglected without actually checking that they are indeed zero.  We shall show in this section that these terms are in fact not  zero for all Poincar\'e transformations and need therefore special treatment.

The difficulty comes from the Lorentz boosts, to which we explicitly turn (the
surface terms for the other Poincar\'e transformations can be easily verified
to raise no problem as it will be explicitly checked in Subsection \ref{sec:EMhamilSym00} below).

The symplectic $2$-form derived from (\ref{eq:StartingPoint}) is
\be
  \Omega = \int d^3x \, d_V \pi^i \, d_V A_ i
 \ee
 where the product is the exterior product $\wedge$ of forms which we are not writing explicitly, and where we use the symbol $d_V$ for the exterior derivative in phase space in order not to introduce confusion with the spacetime exterior derivative.

The transformation defined by the vector field $X$ is canonical if $d_V \left( i_X \Omega \right) = 0$.  Evaluating this expression for the boosts $ \xi = br$, one finds
\be
d_V  \left( i_b \Omega \right) = \int d^3 x \, \d_m \left(\sqrt{g} \xi d_V F^{mi} \right) \, d_V A_i
\ee
where we have used $d_V \pi^i \, d_V \pi_i = 0$.  Integrating by parts and using $d_V F_{ij} \, d_V F^{ij} = 0$, we get that $d_V  \left( i_b \Omega \right)$ reduces to a surface term,
\be
d_V  \left( i_b \Omega \right) = \oint d^2 x \,  \sqrt{g} \, \xi \,  d_V F^{ri} \, d_V A_i \, ,
\ee
an expression that can be transformed to
\begin{eqnarray}
d_V  \left( i_b \Omega \right) &= & - \oint d^2 x  \sqrt{\xbar \gamma} \, b  \, d_V  \xbar D ^A  \xbar A_r \, d_V \xbar A_A  \\
&=& \oint d^2 x  \sqrt{\xbar \gamma}  \, d_V   \xbar A_r \,  \xbar D ^A (b \,  d_V \xbar  A_A) \label{eq:unwanted}
\end{eqnarray}
using the asymptotic form of the fields.  This term would vanish if $\xbar A_A$ was
even,  but in our case, $\xbar A_A$ has a non trivial odd component. That odd
component remains crucially present, and $d_V  \left( i_b \Omega \right)$ does
not vanish, even if it reduces to an improper gauge transformation, $\xbar A_A
= \partial_A \xbar \Phi$ with $\xbar \Phi$ even. 
 Something must thus be done in order to accommodate Lorentz boosts. 

One might try to impose a relationship between $\xbar A_r$ and $\xbar A_A$ so that the $2$-form $$\oint d^2 x  \sqrt{\xbar \gamma} \, b  \, d_V  \xbar D ^A  \xbar A_r \,  d_V \xbar A_A$$ vanishes.  This would be the analog of what was done in \cite{Henneaux:2002wm,Henneaux:2004zi,Henneaux:2006hk} when dealing with a slowly decaying scalar field in anti-de Sitter space.  There is, however, no obvious relationship that can be imposed without at the same time destroying the asymptotic angle-dependent $u(1)$ at infinity,  so that a different route must be followed.

We shall instead add a surface degree of freedom at infinity, which we denote
by $\xbar \Psi$\footnote{The boundary field $\xbar\Psi$ plays here a similar role
	to the extra boundary field $\phi_\pm$ introduced in \cite{He:2014cra} in order to
complete the Poisson structure at future and past null infinity. Both fields parametrise improper
gauge degrees of freedom. Using the results of appendix \ref{app:u1transfos}, one can show
that the fields $\phi_\pm$ defined at null infinity are linear combinations of
$\xbar\Psi$ and $\xbar\Phi$.}.  The field $\xbar \Psi$ is at this stage a field living on the two-sphere at infinity, which can depend on time.  To cancel the unwanted surface term (\ref{eq:unwanted}), we add to the symplectic $2$-form the surface term
\be
- \oint d^2x \, \sqrt{\xbar \gamma} \, d_V \xbar A_r \, d_V \xbar \Psi \label{eq:SympSurfTerm}
\ee
and postulate that $\xbar \Psi$ transforms under boosts as
\be
\delta_b \xbar \Psi = \xbar D^A (b \xbar A_A) + b \xbar A_r. \label{eq:VarPsi}
\ee
The field $\xbar \Psi$ could be restricted to be odd under parity, i.e., to be of the same parity as $\xbar A_r$.  We can however add to it an even part, which is clearly pure gauge since it drops from the surface term (\ref{eq:SympSurfTerm}). Shifting $\xbar \Psi$ by an arbitrary even function is then another proper gauge symmetry of the theory, which comes in addition to the proper gauge symmetries generated by $G[\epsilon]$ with $\epsilon$ odd.  In other words, we take the field $\xbar \Psi$ to be odd up to a proper gauge symmetry. 

The second term in the right-hand side of (\ref{eq:VarPsi}) is even.  It is
therefore a pure gauge term, which we find convenient to add as it simplifies
some formulas below.  (When extended into the bulk, the transformation takes the convenient form (\ref{eq:extensionPsi}).)

We also modify the transformation law of $A_i$ by adding a gauge transformation parametrized by the gauge parameter $\xi \Psi$, where $\Psi$ is any function that matches $\xbar \Psi$ at infinity as
\be
\Psi = \frac{\xbar \Psi}{r} + \frac{\Psi^{(1)}}{r^2} + o(r^{-2}).  \label{eq:BulkPsi}
\ee
Which extension of $\xbar \Psi$ one takes does not matter since two extensions will differ by a proper gauge transformation.  The full transformation of $A_i$ under boosts is thus
\be
\delta A_i = \frac 1 {\sqrt g}\xi \pi_i   + \partial_i (\xi \Psi). \label{eq:VarACorrected}
\ee
The extra gauge transformation does not yield  extra term  in the transformation of $\xbar A_r$ ($\d_r (b \Psi)$ is of order $r^{-2}$), but does induce an extra surface term in the variation of the symplectic form equal to 
\be
 \oint d^2x b d_V \xbar\pi^r d_V \xbar \Psi.
\ee
With these transformations, the symplectic form is invariant and the boosts define canonical transformations.
Note that the even part of $\xbar \Psi$ in (\ref{eq:VarACorrected}) defines a proper gauge transformation, while its odd part defines an improper gauge transformation.

\subsection{Complete Action}
The action describing the dynamics with the new field $\xbar \Psi$  included is
\begin{multline}
\label{eq:ActionNBC}
	S_H[A_i, \pi^i, \xbar \Psi ; A_t] = \int dt \left\{ \int d^3x \, 
		\pi^i \d_t
		A_i 
		 	- \int d^3x \left(   
				\frac{1}{2 \sqrt{g}} \pi^i \pi_i + \frac{1}{4} \sqrt{g} F^{ij} F_{ij}  + A_t  \mathcal{G}\right) \right.\\ 
				\left. - \oint d^2x \, \sqrt{\xbar \gamma} \, \xbar A_r \,  \d_t \xbar \Psi \right\}.
\end{multline}

Variations of the extra boundary term in (\ref{eq:ActionNBC}) give
two extra equations of motion on the boundary:
\begin{equation}
	\d_t \xbar A_r = 0, \qquad \d_t \xbar \Psi = 0.
\end{equation}
The first one is not new as it is a consequence of the bulk equation of motion
generated by $\pi^r$:
\begin{equation}
	\d_t A_r - \d_r A_t - \frac{1}{\sqrt g} \pi_r = 0 \quad \Rightarrow
	\quad \d_t \xbar A_r = 0.
\end{equation}
The second one is an equation of motion for the new surface degree of freedom $\xbar \Psi$.

One could add to the action a surface Hamiltonian involving $\xbar A_r$ that would modify the equation of motion for $\xbar \Psi$.  The choice made here, namely $H_{Surface} = 0$, is the simplest one, and it is compatible with Poincar\'e invariance as discussed below.

\subsection{New global symmetry - Alternative form of the action}
\label{subsec:EMhamilSym}

Note that the gauge generator (\ref{eq:ChargeG0}) remains well-defined even after the symplectic structure has been modified because the leading term of the gauge parameter does not depend on the fields.

It is clear that the action is invariant under arbitrary shifts of $\xbar \Psi$ which can depend on the angles,
\begin{equation}
	\delta_{\mu} \xbar \Psi =  \xbar \mu, \quad \delta_{\mu} A_i = 0, \quad
	\delta_{\mu} \pi^i = 0,  \quad \delta_\mu A_t = 0.
\end{equation}
The even part of the parameter $\mu$, which can have an arbitrary time-dependence,  generates a pure gauge transformation, but the odd part, which must be time-independent in order to leave the action invariant,  generates a global symmetry.  The corresponding charge is  $- \oint d^2x  \sqrt{\xbar \gamma}
	\,\xbar \mu \xbar A_r$ as we shall derive below, and is generically non-vanishing.

It turns out that both the even part of $\xbar \epsilon$ and the odd part of $\xbar \mu$ naturally combine into a single function.  In order to make that structure manifest, it is useful to extend the surface degree of freedom $\xbar \Psi$ into a dynamical bulk field $\Psi$, i.e., to treat (\ref{eq:BulkPsi}) as a field to be varied in the action principle.  This can be done provided one introduces at the same time the constraint that the conjugate $\pi_\Psi$ to $\Psi$ vanishes, 
\be
\pi_\Psi \approx 0
\ee
so that the bulk part of $\Psi$ is pure gauge and only (the odd part of)  its leading term $\xbar \Psi$ in the asymptotic expansion is truly dynamical (\ref{eq:BulkPsi}).  
One can introduce a Lagrange multiplier $\lambda$ for that constraint, so that an alternative action is
\begin{multline}
	S_H[A_i, \pi^i, \Psi, \pi_\Psi; A_t, \lambda] = \int dt \left\{ \int d^3x \, 
		\pi^i \d_t
		A_i + \pi_\Psi \d_t \Psi - \oint d^2x
		\, \sqrt {\xbar\gamma}\,  \xbar A_r \d_t \xbar \Psi
		\right.\\
		\left. 	- \int d^3x \left(\frac{1}{2\sqrt g} \pi^i \pi_i + \frac{\sqrt
		g}{4} F^{ij} F_{ij} \right)  - \int d^3x \left( \lambda \pi_\Psi +  A_t \mathcal{G}  \right)\right\}.
\end{multline}
In order to avoid a logarithmic divergence in the kinetic term $\int d^3 x \pi_\Psi \d_t \Psi$, we impose that the conjugate momentum $\pi_\Psi$ -- a density of weight one --, behaves asymptotically in spherical coordinates as  \be
\pi_\Psi =  \frac{1}{r} \pi^{(1)}_\Psi +
	o(r^{-1}). \label{eq:bcpiPsi}
\ee
This behaviour  does not remove physical solutions since $\pi_\Psi = 0$ on-shell. 

Once the bulk piece is included, the above shifts of $\Psi$ (with the obvious extension in the bulk of the $\xbar \mu$ -transformation, and assuming that $\xbar \mu$ is field-independent) are generated by 
 the canonical generator 
\begin{equation}
	\label{eq:generatorU1largeI0}
	G_{\mu}[A_i, \Psi,  \pi^i, \pi_\Psi] = \int d^3x \mu \pi_\Psi  - \oint d^2x \sqrt{\xbar \gamma}
	\,\xbar \mu \xbar A_r
\end{equation}
where the boundary term insures that it is an allowed functional.  The generator $G_{\mu}$ reduces to the boundary term when the constraint $\pi_\Psi \approx 0$ holds.   This conserved boundary term does not generically vanish for odd $\xbar \mu$'s  and therefore, it indeed defines then a proper gauge transformation in agreement with what was observed before.   Contrary to the usual framework,  the algebraic constraint $\pi_\Psi$ can thus generate improper gauge transformations with non-vanishing charges.   This is because the symplectic form involves a non-trivial surface contribution.  More information on the canonical formalism with surface-term contributions to the symplectic structure is given in Appendix \ref{sec:PoissonStructure}.

The addition of the surface degree of freedom and the companion boundary modification of the symplectic form lead also to the satisfactory situation that the conserved quantities $- \oint d^2x  \sqrt{\xbar \gamma}
	\,\xbar \mu \xbar A_r$ appear as Noether charges for  definite, non trivial symmetries.  Without this surface degree of freedom, the conserved charge $\xbar A_r$ would appear as a conserved quantity that generates no transformation.
  
To summarize, the extended formulation contains two first-class constraints $\pi_\Psi \approx 0 $ and $\mathcal{G} \approx 0$
that generate the two independent
gauge transformations described above:
\begin{equation}
	\delta_{\mu,\epsilon} \Psi = \mu, \quad \delta_{\mu,\epsilon} A_i =  \d_i \epsilon, \quad
	\delta_{\mu,\epsilon} \pi^i = 0, \quad
	\delta_{\mu,\epsilon} \pi_\Psi = 0
\end{equation}
where the gauge parameters have the asymptotic behaviour 
\be
	\mu = \frac 1 r \xbar \mu(x^A) + \frac{1}{r^2} \mu^{(1)}(x^A) +
	o(r^{-2}), 
	\quad \epsilon = \xbar \epsilon(x^A) + \frac{1}{r} \epsilon^{(1)}(x^A) +
	o(r^{-1}).
\ee
The gauge transformations are improper when the leading terms $\xbar \mu$, respectively  $\xbar \epsilon$, have an odd, respectively even, part.   There is thus a wealth of two functions of the angles, one odd and the other even, of improper gauge transformations.  
Assuming that $\xbar \mu$ and $\xbar \epsilon$  are field-independent, we can write the total generator:
\begin{equation}
	\label{eq:generatorU1largeI}
	G_{\mu,\epsilon}[A_i, \Psi,  \pi^i, \pi_\Psi] = \int d^3x (\mu \pi_\Psi + \epsilon
	\mathcal{G}) + \oint d^2x (\xbar \epsilon \xbar \pi^r - \sqrt{\xbar \gamma}
	\,\xbar \mu \xbar A_r)
\end{equation}
and
easily check that it is conserved
\begin{equation}
	\{ G_{\mu, \epsilon}, H\} = \int d^3x \, \pi^i \d_i \mu \approx 0.
\end{equation}

To complete the formulation, we need to give boundary conditions on the  Lagrangian multiplier $\lambda$.  Just as we did for $A_t$, we allow it  to behave as a gauge transformation asymptotically,
\be
\lambda = \frac{\xbar \lambda}{r} + \frac{\lambda^{(1)}}{r^2} + o(r^{-2}). \label{eq:BulkLambda}
\ee
When $ \xbar \lambda(-x^A) =-\xbar \lambda(x^A)$, the gauge transformation is improper.

We have written the action principle assuming that the electric charge $\oint d^2x \xbar \pi^r$ was fixed,  allowing the asymptotic  value of the ``conjugate chemical potential'' $\xbar A_t$ to vary.  Alternatively, one can fix $\xbar A_t$ and vary the electric charge, in which case one needs to add the surface term $\int dt \oint d^2x \xbar A_t \xbar \pi^r$ at infinity.  Similarly we have assumed that the other charge $\oint d^2x \xbar A_r$ was fixed since otherwise one gets a clash between the equations of motion obtained by varying $\pi_\Psi$ ($\d_t \Psi = \lambda$) and $\xbar A_r$ ($\d_t \xbar \Psi = 0$), when $\lambda$ involves a transformation at infinity.  In fact, the leading term $\xbar \lambda$ of the Lagrange multiplier acts as chemical potential for the charge $\oint d^2x \xbar A_r$.   If one adds to the action the surface term $\int dt \oint d^2x \xbar \lambda \, \xbar A_r$, one goes to the  corresponding ``grand canonical ensemble'' and the charge  $\oint d^2x \xbar A_r$ can now be varied while keeping  $\xbar \lambda$ fixed.

Finally, we note that there is no particular simplification if $\xbar A_A =
\partial_A \xbar\Phi$.  The only formula that is affected is
(\ref{eq:VarPsi}), which becomes $ \delta_b \xbar \Psi = \xbar D^A (b \d_A
\xbar\Phi) + b \xbar A_r$.

\subsection{Poincar\'e charges}
\label{sec:EMhamilSym00}

With the boundary modification of the symplectic charges, all Poincar\'e
transformations are canonical transformations with a well-defined generator.  We first consider the kinematical transformations, i.e., spatial translations and rotations.

As we already stressed before,   transformations of the fields under a symmetry are defined up to a gauge transformation in any gauge theory.  For spatial translations and rotations, we adjust the gauge transformation in such a way that the action of these spatial symmetries on the fields is the ordinary Lie derivative, i.e.,
\be
\delta_{\xi^k} A_i = 
  \cL_{\xi^k} A_i, \quad  \delta_{\xi^k}\Psi =  \cL_{\xi^k} \Psi 
  \ee
 where $\Psi$ is a spatial scalar so that $\cL_\xi \Psi = \xi^k \d_k \Psi$ and where the spatial vector $\xi^k$ is given by (\ref{eq:FormOfW}).
 This choice, which is somewhat arbitrary,   leads to a simple algebra of the charges.  Spatial translations and rotations are then generated by
\begin{flalign}
	\label{eq:rotgen}
	P_{0,\xi^i} & = \int d^3x( \pi^i \cL_{\xi^k} A_i + \pi_\Psi \cL_{\xi^k} \Psi) 
	+ \oint d^2x \sqrt{\xbar\gamma} \,\xbar \Psi Y^A\d_A\xbar A_r\\
	& = \int d^3x\,\xi^i( F_{ij} \pi^j  - A_i \d_j \pi^j+ \pi_\Psi \d_i \Psi) 
	+ \oint d^2x\, Y^A\left(\xbar A_A\xbar \pi^r + \sqrt{\xbar\gamma} \,\xbar \Psi
		\d_A\xbar A_r\right).
\end{flalign}
The boundary term in the expression of the rotation charges (angular momentum) is essential with the twisted boundary conditions in order to have an allowed functional.

The generator of the boosts and 
time translation is more complicated.  We have worked out in Subsection
\ref{subsec:BoostsSymp} the transformation rules of $A_i$ and $\xbar\Psi$
under boosts, equations (\ref{eq:VarACorrected}) and (\ref{eq:VarPsi}),
respectively. Noting that $\xbar D^A (b A_A) + b \xbar A_r$ is the leading
term in the asymptotic value of $\nabla^i (\xi A_i)$ with $\xi= br$, we can
extend the transformation law of $\xbar\Psi$ into the bulk with
\begin{equation}
	\delta_b \Psi = \nabla^i(\xi A_i) .  \label{eq:extensionPsi}
\end{equation}
 The boosts are then generated as
\begin{multline}
	\label{eq:boostgen}
	P_{\xi,0}= \int d^3x\, \xi \left(-\Psi \d_i\pi^i -
	 A_i \nabla^i\pi_\Psi + \frac{1}{2\sqrt g} \pi_i \pi^i +
 \frac{\sqrt g}{4} F_{ij} F^{ij} \right)\\
	+ \oint d^2x \,b \Big(\xbar \Psi \xbar \pi^r
		 +\sqrt {\xbar\gamma}  \xbar
	A_A \xbar D^A \xbar A_r\Big).
\end{multline}
We take the same formal expression for the generator of time translations where the parameter $\xi$ is given by $\xi = T$ with $\d_A T=0$.  This amounts again to a specific choice of the improper gauge transformation included in what is meant by a ``time translation'' and is again a matter of choice.  Our choice leads to a simple algebra.  The generator (\ref{eq:boostgen}) is thus generally valid for
\begin{equation}
	\xi = b(x^A)r + T, \quad D_AD_B b + \xbar\gamma_{AB}b=0,
	\quad \d_A T = 0.
\end{equation}
While the surface term is mandatory for the boosts in order to have a well-defined generator,  it vanishes for time translations.   
It should be stressed that the logarithmic divergence in the bulk integral of $P_{\xi,\xi^i}$, potentially present for rotations and boosts, cancels due to our asymptotic fall-off.  

The total generator of Poincar\'e transformations can be written in terms of local diffeomorphisms
generators in the following way
\begin{gather}
	P_{\xi, \xi^i} = \int d^3x \left( \xi \mathcal H^{EM} + \xi^i
	\mathcal H^{EM}_i\right) + \mathcal B^{EM}_{(\xi, \xi^i)}, \label{eq:PoincEM26}\\
	\mathcal H^{EM} = -\Psi \d_i\pi^i -
	 A_i \nabla^i\pi_\Psi + 
	 \frac{1}{2\sqrt g} \pi_i \pi^i +
	 \frac{\sqrt g}{4} F_{ij} F^{ij},\\
	  \quad \mathcal H^{EM}_i = F_{ij} \pi^j - A_i \d_j \pi^j+ \pi_\Psi
	 \d_i \Psi,\\
	\mathcal B^{EM}_{\xi, \xi^i} = \oint d^2x \left(b (\xbar \Psi \xbar \pi^r
		 +\sqrt {\xbar\gamma}  \xbar
		A_A \xbar D^A A_r) +  Y^A(\xbar A_A\xbar \pi^r +
		\sqrt{\xbar\gamma} \,\xbar \Psi
	\d_A\xbar A_r)\right).
\end{gather}

\vspace{5mm}

One can easily compute the algebra of the various generators.  One finds:
\begin{gather}
	\label{eq:EMalgebraI}
	\{P_{\xi_1, \xi^i_1},P_{\xi_2, \xi^i_2}\} = P_{\hat\xi, \hat
	\xi^i}, \\
	\{G_{\mu,\epsilon}, P_{\xi,\xi^i}\} = G_{\hat \mu, \hat
	\epsilon},\quad
	\{G_{\mu_1,\epsilon_1}, G_{\mu_2,\epsilon_2}\} = 0,\\
	\hat \xi = \xi_1^i \d_i \xi_2 - \xi_2^i \d_i \xi_1,
	\label{eq:EMalgebraII}
	\quad \hat \xi^i = \xi_1^j \d_j \xi_2^i - \xi_2^j \d_j \xi_1^i +
	g^{ij}(\xi_1 \d_j \xi_2 - \xi_1 \d_j\xi_2),\\
	\label{eq:EMalgebraIII}
	\hat \mu = \nabla^i(\xi\d_i \epsilon)-\xi^i\d_i \mu,\quad
	\hat\epsilon = \xi \mu-\xi^i\d_i \epsilon. 
\end{gather}
It follows from these equations that the algebra of the symmetries is a
semi-direct sum of the Poincar\'e algebra and the abelian algebra parametrized by $\xbar\mu$
and $\xbar\epsilon$. The action of the Poincar\'e subalgebra characterising
this semi-direct sum is easily obtained from \eqref{eq:EMalgebraIII}:
\begin{gather}
	\delta_{(Y,b,T,W)}\xbar \mu = Y^A\d_A \xbar \mu - \xbar D_A(b \xbar D^A\xbar
	\epsilon), \quad \delta_{(Y,b,T,W)}\xbar \epsilon = Y^A \d_A \xbar
	\epsilon - b \xbar \mu.  \label{eq:EMalgebraXX}
\end{gather}
If we had chosen a different improper gauge transformation to accompany the
spacetime transformations, the bracket $\{P_{\xi_1, \xi^i_1},P_{\xi_2,
	\xi^i_2}\} \approx P_{\hat\xi, \hat
	\xi^i}$ would have generically been modified by terms containing $G$.  
	
	In Appendix \ref{sec:EMhyperbolic}, we relate our study to
the Lagrangian analysis in hyperbolic slicings, and in Appendix \ref{app:u1transfos},  we prove that the above algebra agrees with the one obtained at null infinity.  In particular, we show how the even and odd gauge functions $\epsilon$ and $\mu$ combine to form the full angle-dependent $u(1)$ transformations seen at null infinity.  This analysis is very close to the corresponding one for gravity \cite{Troessaert:2017jcm}, \cite{Henneaux:2018cst}.

\section{Asymptotic analysis:  II. Magnetic charges}
\label{sec:Magnetic}

The twisted parity conditions that we just investigated lead to a perfectly consistent picture and a well-defined Hamiltonian formulation.  They suffer, however, from two drawbacks.  First, they do not allow for the presence of magnetic sources.  Second, they generically lead to divergences in the magnetic field as one approaches either the past of future null infinity or the future of past null infinity.  This is shown in Appendix \ref{sec:EMhyperbolic}.   To avoid these singularities, one must impose the extra condition that the even parts of the angular components of the electric field and of $F_{AB}$ vanish.  

We shall for these reasons turn to the boundary conditions (\ref{eq:BCMagn1})
and (\ref{eq:BCMagn2}) which do not suffer from these drawbacks: they cover
magnetic sources as we indicated previously, and are free from singularities
as one tends to null infinity.

It turns out that the asymptotic analysis of these boundary conditions is straightforward.  The reason is that  the modification of the symplectic term and the addition of non-trivial surface terms to the Poincar\'e generators is forced by the odd part $\partial_A \Phi$ of the angular component $A_A$ of the vector potential.  The even part drops out in those considerations and raises no extra complication.  All the difficulties have thus been already treated in the previous section.

The action takes thus again the form
\begin{multline}
	S_H[A_i, \pi^i, \Psi, \pi_\Psi; A_t, \lambda] = \int dt \left\{ \int d^3x \, 
		\pi^i \d_t
		A_i + \pi_\Psi \d_t \Psi - \oint d^2x
		\, \sqrt {\xbar\gamma}\,  \xbar A_r \d_t \xbar \Psi
		\right.\\
		\left. 	- \int d^3x \left(\frac{1}{2\sqrt g} \pi^i \pi_i + \frac{\sqrt
		g}{4} F^{ij} F_{ij} \right)  - \int d^3x \left( \lambda \pi_\Psi +  A_t \mathcal{G}  \right)\right\}
\end{multline}
and the Poincar\'e generators are
\begin{gather}
	P_{\xi, \xi^i} = \int d^3x \left( \xi \mathcal H^{EM} + \xi^i
	\mathcal H^{EM}_i\right) + \mathcal B^{EM}_{(\xi, \xi^i)}, \label{eq:PoincEM50}\\
	\mathcal H^{EM} = -\Psi \d_i\pi^i -
	 A_i \nabla^i\pi_\Psi + 
	 \frac{1}{2\sqrt g} \pi_i \pi^i +
	 \frac{\sqrt g}{4} F_{ij} F^{ij},\\
	  \quad \mathcal H^{EM}_i = F_{ij} \pi^j - A_i \d_j \pi^j+ \pi_\Psi
	 \d_i \Psi,\\
	\mathcal B^{EM}_{\xi, \xi^i} = \oint d^2x \left(b (\xbar \Psi \xbar \pi^r
		 +\sqrt {\xbar\gamma}  \xbar
		A_A \xbar D^A A_r) +  Y^A(\xbar A_A\xbar \pi^r +
		\sqrt{\xbar\gamma} \,\xbar \Psi
	\d_A\xbar A_r)\right).
\end{gather}
The logarithmic divergences in the bulk integrals are absent because the
leading $O(r^{-1})$-term in the integrand of the boost and rotation generators
is odd and thus integrates to zero over the 2-sphere.  One may also replace
$\xbar A_A$ by $\partial_A \xbar \Phi$ in the boundary terms, because $\xbar
A_A^{\hbox{\footnotesize{even}}}$ simply drops out. All the subtle features
are introduced by the odd component of $\xbar A_A$ studied above.

The theory possesses the same improper gauge transformations as above generated by
\begin{equation}
	\label{eq:generatorU1largeIV}
	G_{\mu,\epsilon}[A_i, \Psi,  \pi^i, \pi_\Psi] = \int d^3x (\mu \pi_\Psi + \epsilon
	\mathcal{G}) + \oint d^2x (\xbar \epsilon \xbar \pi^r - \sqrt{\xbar \gamma}
	\,\xbar \mu \xbar A_r).
\end{equation}
These transformations act only on the  $\partial_A \xbar \Phi$-part of $\xbar
A_A$ and on $\xbar \psi$ so that here again, all the conceptual points have been made in the previous section. 

This ends the discussion of the general case.

\section{Conclusions}
\label{sec:Conclusions}

In this paper, we have proposed new boundary conditions for electromagnetism at spatial infinity.  These are
(\ref{eq:hamilasymptI0}), (\ref{eq:hamilasymptI022}), (\ref{eq:parityRadial}), (\ref{eq:BCMagn1}),  (\ref{eq:BCMagn2}), (\ref{eq:BulkPsi}) and (\ref{eq:bcpiPsi}).   These boundary conditions include magnetic sources and fulfill the consistency requirements listed in the introduction.  They are invariant under an infinite set of non trivial asymptotic symmetries, labeled by one function on the two-sphere. These angle-dependent $u(1)$ transformations have generically non-vanishing charges and their generator is given by (\ref{eq:generatorU1largeIV}). The way the even component ($\epsilon$) and the odd component ($\mu$) of the symmetry parameters combine as compared with the analysis at null infinity is somewhat subtle and worked out in Appendix \ref{app:u1transfos}.   We have also worked out the canonical generators of the Poincar\'e group, which involves an extra surface degree of freedom (Eq. (\ref{eq:PoincEM50}), and computed their algebra with the asymptotic infinite angle-dependent $u(1)$ symmetry (Eq. (\ref{eq:EMalgebraI})-(\ref{eq:EMalgebraIII})).   Note that if the charge is not zero, Lorentz invariance is broken since the angle-dependent conserved quantity $G_{\mu,\epsilon}$ transforms non trivially under rotations and boosts, as clearly exhibited by (\ref{eq:EMalgebraXX}).

A crucial element in the emergence of the non trivial symmetry at spatial infinity is the twist in the parity conditions on the angular components of the vector potential as compared with the conditions of \cite{Henneaux:1999ct}.  This twist is the electromagnetic analog of what was done in \cite{Henneaux:2018cst} for pure gravity.  However, as we have shown here, it is sufficient for the twist to be given by an improper gauge transformation.  This restricted class of twists enables one to include magnetic monopoles. The same can be done for gravity in order to encompass the Taub-NUT solution, as it will be shown elsewhere.

We stress that the parity conditions given here on the angular components of the potential are not a restriction in the sense that they do not exclude the standard solutions relevant for the usual asymptotic analysis.   If they were imposed on the next order, they might of course be a restriction, but they are here conditions on an order that is actually zero for the Li\'enard-Wichiert potential that plays such an important role in the analysis of the fields at null infinity.  We have in fact shown that our results are in agreement with those obtained at null infinity.  The non trivial symmetries visible at null infinity do have an expression at spatial infinity and turn out to be also asymptotic symmetries with charges that reduce to surface terms at null infinity (``improper gauge transformations''), with no bulk contribution.  Relaxing the parity conditions does not seem necessary from the point of view of known solutions and furthermore, could lead to undesirable divergences in the symplectic structure.

Our work also shows an interesting interplay between the radial components of the fields and their angular components.  The radial components, associated with the Coulomb aspect of the electromagnetic field, are subject to parity conditions that reflect what is left of spherical symmetry in a moving frame. At the same time, they act as generators of the infinite-dimensional symmetry on the angular components and dictate what is a proper gauge symmetry and what is an improper one.  To allow for the improper ones, the angular components must have a parity-twisted piece with respect to the radial components. 

There is a clear difference between 
 electromagnetism and
gravity in the asymptotic treatment given here. While the twisted parity conditions are sufficient in the gravity case to ensure a consistent formulation without modification of the symplectic structure, an extra feature appears in the case of electromagnetism.  It is that the symplectic structure must be modified by a boundary contribution (as observed first in \cite{Campiglia:2017mua}), and that surface degrees of freedom must be explicitly introduced in a gauge-fixing free approach.  It is thanks to this extra structure that the full angle-dependent algebra (and not just that of the even functions) emerge at infinity.

In the case of gravity, the condition that the mixed component $h_{rA}$ of the spatial metric should vanish to leading order was imposed in \cite{Henneaux:2018cst}.  If it is not satisfied, the boost charges are not integrable.  The treatment of the electromagnetic field given here suggests that one might avoid the condition $\xbar h_{rA}= 0$  in gravity by introducing surface degrees of freedom (which could be related to the lapse and the shift by gauge-fixing) and modifying the symplectic structure by a surface term.  This might lead to an enlargement of the asymptotic symmetry algebra.  We recall in this context that super-rotations \cite{Banks:2003vp,Barnich:2009se,Barnich:2010eb,Barnich:2011mi,Barnich:2016lyg}
were not included in the approach of  \cite{Henneaux:2018cst}. Perhaps this might open the way to their satisfactory inclusion.  
 Similarly it would be interesting to extend the analysis to the Yang-Mills case, as well as to supergravity where one would have to work out the impact of the twist in parity conditions on the spinors.   It is hoped to return to all these questions in the future.

\section*{Acknowledgments} 
MH is grateful to the Institute for Advanced Study (Princeton) for kind hospitality while this work was completed.  This research was partially supported by the ERC Advanced Grant ``High-Spin-Grav'', by FNRS-Belgium (convention FRFC PDR T.1025.14 and  convention IISN 4.4503.15) and by the ``Communaut\'e Fran\c{c}aise de Belgique'' through the ARC program.

\begin{appendix}

\section{Poisson structure}
\label{sec:PoissonStructure}

We extend in this appendix the work of  \cite{Regge:1974zd} and
 \cite{Brown:1986ed} to the case when the symplectic form is not the standard one but is modified by a boundary term.  This extension relies on well-established results of the hamiltonian formalism, as described e.g. in the monograph \cite{Arnold}.

In the following, we
will collectively denote the fields by $z^A$ and we will 
restrict our analysis to symplectic structures $\Omega$ of the form
\begin{equation}
	\Omega[z;d_Vz, d_Vz] = \int \frac{1}{2}\sigma_{AB} d_V z^A d_V
	z^B\, d^nx + \omega[z;d_Vz, d_Vz]
\end{equation}
where $\omega$ is a boundary term such that $d_V \omega = 0$. We will also assume
the matrix $\sigma_{AB}=-\sigma_{BA}$ to be constant and invertible
$\sigma^{AB} \sigma_{BC} = \delta^A_C$.

Let us consider a vertical vector 
field $\delta_F z^A$ preserving the
asymptotic conditions. It is a hamiltonian
vector field iff
\begin{equation}
	\label{eq:defallowed}
	F[z^A] = \int f(z) \, d^nx + \oint f_b(z) \, d^{n-1} x \quad \text{s.t.}
	\quad d_V F = -i_F \Omega[z;d_Vz, d_Vz]
\end{equation}
where $\Omega$ is a closed vertical 2-form. 
We will call allowed functionals $F$ functionals that are associated to a
hamiltonian vector through \eqref{eq:defallowed}. As in the case of
Regge-Teitelboim, this prescription fixes the boundary term of allowed
functionals up to a constant.

Considering two allowed functionals $F_1$ and $F_2$, we can define their
Poisson bracket as
\begin{equation}
	\{ F_1, F_2\} = i_1 i_2 \Omega = \int \frac{\delta F_1}{\delta
	z^A} \sigma^{AB} \frac{\delta F_2}{\delta z^B} d^nx + i_1 i_2 \omega.
\end{equation}
In this case, we see that the usual result given by the bulk term
is modified by a boundary term. By construction, the Poisson bracket is 
antisymmetric and we have the following property
\begin{equation}
	\{ F_1, F_2\} = -i_2i_1\Omega = i_2 d_V F_1 = \delta_2 F_1
\end{equation}
where $\delta_2 z^A$ is the variation associated to $F_2$.
Using standard techniques, we can then prove the two following results that make this Poisson bracket
well-defined.
\begin{theorem}
	The Poisson bracket of two allowed functionals $F_1$ and $F_2$ is 
	an allowed functional with
	\begin{equation}
		\label{eq:theointernal}
		d_V\{F_1, F_2\} = -i_{[2,1]} \Omega.
	\end{equation}
\end{theorem}
	
\begin{theorem}
	The Poisson bracket satisfies Jacobi identity:
	\begin{gather}
		\{\{F_1, F_2\}, F_3\} + cyclic = 0.
	\end{gather}
\end{theorem}
The demonstration of these theorems follow usual lines of symplectic geometry and is left to the reader.
	
In the
particular case where the boundary term of the symplectic structure is 
absent,  $\omega = 0$, the allowed functionals $F$ are those 
with a variation of the form
\begin{equation}
	d_V F = \int \, \frac{\delta F}{\delta z^A} d_V z^A d^nx - i_F \omega
	= \int \, \frac{\delta F}{\delta z^A} d_V z^A d^nx
\end{equation}
without boundary term.   This is the  prescription of \cite{Regge:1974zd}.

\section{Electromagnetism in hyperbolic slicings}
\label{sec:EMhyperbolic}

\subsection{Connection with previous work}

The connection between spatial infinity and null infinity is most conveniently worked out in hyperbolic coordinates \cite{Ashtekar:1978zz,BeigSchmidt,Beig:1983sw,Ashtekar:1991vb},
\begin{gather}
	\label{eq:minkHyperbol}
	g_{\mu\nu} dx^\mu dx^\nu = d\eta^2 + \eta^2 h_{ab} dx^adx^b, \\
	h_{ab} dx^adx^b = \frac{-1}{(1-s^2)^2}ds^2 + \frac{1}{1-s^2}  \xbar
	\gamma_{AB}dx^Adx^B
\end{gather}
where $\xbar \gamma_{AB}$ is the metric on the 2-sphere. The radial coordinate
$\eta$ and the hyperbolic time $s$ are related to the usual time $t$ and
equal time radial distance $r$ by
\begin{equation}
	\eta = \sqrt{-t^2 + r^2}, \qquad s = \frac t r.
\end{equation}
These coordinates only cover the part of space-time where $r\ge \vert t\vert$
which implies $-1 \le s \le 1$. Tensors appearing in the $\eta$ expansions are
defined on the 3 dimensional unit hyperboloid $\mathcal H$ with metric
$h_{ab}$ coordinates $x^a=(s,x^A)$ and covariant derivative  $\mathcal D_a$.   A translation in the hyperbolic time $s$ involves asymptotic boosts at infinity. 

The analysis of electromagnetism in hyperbolic coordinates was performed in
\cite{Campiglia:2017mua}, in which the need to modify the symplectic form by a
boundary term was first pointed out.  In this appendix, we compare our
findings in standard coordinates with the results of \cite{Campiglia:2017mua}.
In the next appendix, we use the hyperbolic coordinates to relate the
asymptotic symmetry group at spatial infinity with the asymptotic symmetry
group at null infinity.

Reference \cite{Campiglia:2017mua} assumes the following fall-offs
for the electromagnetic potential:
\begin{equation}
	\label{eq:EMasympHyperbol}
	A_a = \xbar A_a + A_a^{(1)} \frac{1}{\eta} + o(\eta^{-1}),
	\qquad
	A_\eta = \xbar A_\eta \frac{1}{\eta} + A_\eta^{(1)} \frac{1}{\eta^2}
	+ o(\eta^{-2}).
\end{equation}
With this asymptotic behaviour, the usual bulk action of electromagnetism  is not well-defined. A general variation of the action leads indeed
to:
\begin{multline}
	\label{eq:EMvarAction}
	\delta \int d^4x \frac{-\sqrt{-g}}{4} F_{\mu\nu} F^{\mu\nu} = 
	\int d^4x \sqrt{-g} \nabla_\mu F^{\mu\nu} \delta A_\nu\\
	+ \left.\int d\eta \oint d^2x \sqrt{-g} F^{s\nu} \delta
	A_\nu \right\vert_{s_0}^{s_1} +
	\int_{\mathcal H} d^3x \sqrt{-h} \mathcal D^a \xbar A_\eta \delta
	\xbar A_a
\end{multline}
where $\nabla_\mu$ is the covariant derivative associated to the background
metric \eqref{eq:minkHyperbol}. There is a boundary term at the spatial boundary ${\mathcal H}$.

A solution for removing this unwanted term that allows one to keep a
non-trivial asymptotic symmetry algebra without having an impact on the
solution space is given in \cite{Campiglia:2017mua}.
One adds the following boundary term to the action:
\begin{equation}
	\label{eq:EMactionHyperbol}
	S[A_\mu] = \int d^4x \frac{-\sqrt{-g}}{4} F_{\mu\nu} F^{\mu\nu} +
	\int_{\mathcal H} d^3x \sqrt{-h} \, \xbar A_\eta (\mathcal D^a \xbar A_a
	+ \xbar A_\eta).
\end{equation}
Here, we have used a boundary term quadratic in
$\xbar A_\eta$ but a more general function of $\xbar A_\eta$ could have been
used.

A variation of the action now gives
\begin{multline}
	\delta S = 
	\int d^4x \sqrt{-g} \nabla_\mu F^{\mu\nu} \delta A_\nu +
	\int_{\mathcal H} d^3x \sqrt{-h} \Big(\mathcal D^a \xbar A_a + 2\xbar
	A_\eta\Big) \delta
	\xbar A_\eta\\
	+ \left[\int d\eta \oint d^2x \sqrt{-g} F^{s\nu} \delta
	A_\nu  + \oint d^2x \sqrt{-h} h^{sa} \xbar A_\eta \delta \xbar
A_a\right]_{s_0}^{s_1}.
\end{multline}
The choice of boundary term added to the action leads to an extra equation of motion on the
boundary:
\begin{equation}
	\label{eq:appextraEOM}
	\mathcal D^a \xbar A_a + 2 \xbar A_\eta = 0.
\end{equation}
This is the leading term of Lorenz gauge condition:
\begin{equation}
	\label{eq:appextraEOM2}
	\nabla^\mu A_\mu = \frac{1}{\eta^2} (\mathcal D^a \xbar A_a + 2 \xbar
	A_\eta) + o(\eta^{-2}).
\end{equation}
The extra EOM \eqref{eq:appextraEOM} is compatible with all bulk EOM and 
can be interpreted as an
asymptotic gauge fixing. This condition is a central element of the
analysis of \cite{Campiglia:2017mua} where it is imposed as an extra asymptotic
condition (see also \cite{Afshar:2018apx}). We have just shown that it consistently follows in fact dynamically
from the action. The gauge transformations preserving the action are generated
by gauge parameter of the form
\begin{equation}
	\lambda(\eta, x^a) = \xbar \lambda(x^a) + \lambda^{(1)}(x^a) \frac 1
	\eta + o(\eta^{-1}), \qquad \mathcal D_a \mathcal D^a \xbar \lambda =
	0.  \label{eq:GaugeTransfApp}
\end{equation}
Here and in the next appendix, we used the notation $\lambda$ for the gauge parameter in order to make
the link with \cite{Campiglia:2017mua}. It is not related to the extra
Lagrange multiplier we added in section \ref{subsec:EMhamilSym}.

Our analysis of the gauge symmetries and associated charges, translated in hyperbolic coordinates, reduces to that of \cite{Campiglia:2017mua} if one imposes the gauge condition $A_t = \Psi$, which is admissible if the constant part in the asymptotic expansion of $A_t$ is absent (otherwise, one must impose $A_t - \xbar A_t = \Psi$).  Since the Lagrange multiplier $A_t$ transforms with the time derivative of the gauge parameter $\epsilon$, this gauge condition relates $\d_t \epsilon$ to $\mu$.  Note also that $s$-translations in hyperbolic coordinates involve asymptotic boosts.  It does therefore not come as a surprise that the symplectic form problem encountered in our treatment when handling Lorentz boosts appear here already at the level of ($s$-)time translations.

The action \eqref{eq:EMactionHyperbol} suffers from a second problem: the
associated symplectic structure $\Omega$ is divergent. This problem is solved
in the same way as in the core of our paper by adding parity conditions on
gauge invariant combinations of the asymptotic fields. The discussion proceeds
along parallel tracks and will thus not be repeated here.  

\subsection{Asymptotic behaviour of the fields as one goes to null infinity}

To end this appendix, we will make the link between the asymptotic
fields in hyperbolic coordinates and physical quantities usually defined at
null infinity. The equations of motion derived from action
\eqref{eq:EMactionHyperbol} imply
\begin{equation}
	-(1-s^2) \d_s^2 \xbar A_\eta + \xbar D_A \xbar D^A \xbar
	A_\eta = 0,\quad
	\xbar D^B \xbar F_{BA} - \d_s\left( (1-s^2) \xbar F_{sA}\right) = 0
\end{equation}
where $\xbar F_{ab} = \d_a \xbar A_b - \d_b \xbar A_a$ is the leading term of
the field strength $F_{ab} = \xbar F_{ab} + O(\eta^{-1})$.
These two equations control the asymptotic behaviour of the radial part of 
the electric and magnetic field:
\begin{flalign}
	E^\eta(\eta,x^a) &= \sqrt{-g} F^{s\eta} = \xbar E^\eta(x^a) +
	O(\eta^{-1}), \qquad \xbar E^\eta = -\sqrt{\xbar\gamma} \d_s \xbar
	A_\eta,\\ 
	B^\eta(\eta,x^a) &= \epsilon^{s\eta AB} F_{AB} = \xbar B^\eta(x^a) +
	O(\eta^{-1}), \qquad \xbar B^\eta = - \epsilon^{AB} \xbar F_{AB}.
\end{flalign}
We have denoted by $\epsilon^{\alpha\beta\gamma\delta}$ and $\epsilon^{AB}$ 
the anti-symmetric densities with $\epsilon^{s\eta \theta\phi} = -1 =
-\epsilon^{\theta\phi}$. Using Bianchi's identity, one can easily show that
the leading terms of both $E^{\eta}$ and $B^\eta$ satisfy the following
equation:
\begin{equation}
	- \d_s \left( (1-s^2) \d_s \Xi\right) + \xbar D_A \xbar D^A
	\Xi = 0, \qquad \Xi = \xbar E^\eta, \xbar B^\eta.
\end{equation}
Using spherical harmonics, its general solution can be written in terms of
Legendre polynomials $P_l(s)$ and Legendre functions of the second kind
$Q_l(s)$ which can be expressed in terms of  $P_l(s)$ as
\begin{equation}
	\label{eq:appdefQl}
	Q_l(s) = P_l(s) \frac 1 2 \log \left (\frac{1+s}{1-s}\right) + \tilde
	Q_l(s)
\end{equation}
where $\tilde Q_l(s)$ are polynomials. Without taking into account the parity
conditions, the
corresponding solutions are given as follows
\begin{gather}
	\label{eq:appsolEB}
	\xbar E^\eta(s, x^A) = \sqrt\gamma\sum_{l,m}\left(E^P_{l,m}
	P_l(s) + E^Q_{l,m}
	Q_l(s)\right)
	Y_{l,m}(x^A), \\
	\xbar B^\eta(s, x^A) = \sqrt\gamma\sum_{l,m}\left(B^P_{l,m}
P_l(s) + B^Q_{l,m} Q_l(s)\right) Y_{l,m}(x^A).
\end{gather}
Encoding the contribution from electric sources, $\xbar E^\eta$ is always even
on the hyperboloid which means that $E^Q_{l,m} = 0$. This implies that the
leading part of the
electric field $\xbar E^\eta$ is bounded in the interval $s\in [-1,1]$.
Its magnetic counterpart $\xbar B^\eta$ will also be bounded if one imposes
parity conditions allowing for magnetic charges, $\xbar B^\eta_{odd} = 0 =
B^Q_{l,m}$, 
however, it will diverges at $s=\pm 1$ if one
imposes twisted boundary conditions, $\xbar B^\eta_{even} = 0 = B^P_{l,m}$.

As one could expect, the late time behaviour of both quantities is related to
their value at null infinity.  Unfortunately, the link is not direct as the
hyperbolic coordinates used above are not adapted to the description
of null infinity. We will instead rely on the methods of
\cite{Fried1,Friedrich:1999wk,Friedrich:1999ax} and introduce a new radial
coordinate $\rho = \eta (1-s^2)^{\frac1 2}$. With this, the background metric
takes the form
\begin{equation}
	ds^2 = \frac 1 {(1-s^2)^2} \left ( (1-s^2) d\rho^2 + 2 s \rho ds d\rho
	- \rho^2 ds^2 + \rho^2 \gamma_{AB} dx^Adx^B\right).
\end{equation}
In these new coordinates, the rescaled line element $(1-s^2)^2 ds^2$ is
continuous in the limits $s\to \pm 1$. These two hypersurfaces describe past
and future null infinity with induced coordinates $(\rho, x^A)$. The limits
relevant for Strominger's analysis, namely approaching spatial infinity along
future or past null infinity, correspond to taking the following two limits in
succession: $s\to 1$ or $s\to -1$ and then $\rho\to\infty$.  As the change of
coordinates only involves a rescaling of the radial coordinate, the expansion
of the fields given in \eqref{eq:EMasympHyperbol} behaves nicely. In
particular, the leading term of the radial electric and magnetic fields are
unchanged
\begin{equation}
	E^{\rho}(\rho,x^a) = \sqrt{-g} F^{s\rho} = \xbar E^\eta(x^a) + O(\rho^{-1}), \quad
	B^{\rho} (\rho,x^a)= \epsilon^{s\rho AB} F_{AB} = \xbar B^\eta(x^a) + O(\rho^{-1}).
\end{equation}
Using the explicit solutions obtained in \eqref{eq:appsolEB}, we can now
evaluate the value of the electromagnetic field at null infinity.  The leading
part of the electric field $E^\rho$ is finite and its parity properties imply
that it will satisfy the antipodal matching conditions of Strominger
\cite{Strominger:2013lka}. However, if one imposes twisted boundary
conditions, the magnetic field will in general diverge
in the limits $s\to \pm 1$ which breaks the usual asymptotic conditions of
electromagnetism at null infinity.

\section{Explicit large gauge transformations and link with null infinity}
\label{app:u1transfos}

We compare in this Appendix the asymptotic symmetry group appearing here at spatial
infinity with the asymptotic symmetry group emerging at null infinity.
Using hyperbolic coordinates to describe spatial infinity, large $U(1)$ gauge transformations are parametrized by gauge
parameters of the form
\begin{equation}
	\lambda(\eta, x^a) = \xbar \lambda(x^a) + \lambda^{(1)}(x^a)
	\eta^{-1} + o(\eta^{-1})
\end{equation}
where $\xbar \lambda$ is an even function defined on the
hyperboloid and satisfying
\begin{equation}
	\label{eq:appeqLegendrep}
	\mathcal D_a \mathcal D^a \xbar \lambda = -(1-s^2)^2 \d_s^2 \xbar
	\lambda +
	(1-s^2) \xbar D_A \xbar D^A \xbar \lambda = 0
\end{equation}
(see (\ref{eq:GaugeTransfApp})). On can check that the odd part of $\xbar\lambda$ is always a proper gauge
transformation due to the parity conditions imposed on $\xbar E^\eta$.
The general solution to this equation that is also an even function can be written 
in terms of spherical harmonics as follows
\begin{equation}
	\label{eq:gensoleps}
	\xbar \lambda(s, x^A) = \sum_{l,m} \xbar\lambda_{l,m} \alpha_l(s)
	Y_{l,m}(x^A), \quad \alpha_l = (1-s^2) \d_s Q_l(s)
\end{equation}
where the functions $Q_l(s)$ are Legendre functions of the second kind given
in terms of Legendre polynomials $P_l(s)$ in equation \eqref{eq:appdefQl}. The
fact that $P_l(\pm 1) = (-1)^l$ can be used to determine the asymptotic
behaviour of the functions $\alpha_l$
\begin{equation}
	\lim_{s\to \pm 1} \alpha_l(s) = \lim_{s\to \pm 1} P_l(s) = (\pm 1)^l,
	\qquad \lim_{s\to \pm 1} (1-s^2) \d_s \alpha_l(s) = 0.
\end{equation}
The general solution obtained in \eqref{eq:gensoleps} is completely
characterised by its value at $s=\pm 1$: 
\begin{equation}
	\lim_{s\to\pm 1} \xbar\lambda(s,x^A) =
\epsilon^{\mathcal J}(\pm x^A). 
\end{equation}
These two values correspond to the gauge parameters at future and past null
infinity. As explained in the previous appendix, in order to show this, we
need to introduce a new radial coordinate $\rho = \eta (1-s^2)^{-\frac1 2}$.
The gauge parameters induced at future and past null infinity can then be 
evaluated easily
\begin{equation}
	\lambda(\rho,x^A) \, \vert_{\mathcal J^\pm} = \lim_{s\to\pm 1}
	\lambda(\rho,x^a) =
	\epsilon^{\mathcal J}(\pm x^A) + O(\rho^{-1})
\end{equation}
where sub-leading terms in $\rho$ are present because  the
gauge in the bulk is not fixed. One can see that the antipodal properties of the gauge
parameter obtained in \cite{Strominger:2013lka} are consequences of the parity
conditions associated with the extra equation of motion
\eqref{eq:appextraEOM}. 

The link with the parametrization of the $U(1)$ transformations used in the
Hamiltonian formalism in section \ref{sec:EMhamilSym} is made by parametrising
the function $\xbar \lambda$, solution of equation
\eqref{eq:appeqLegendrep}, in terms of initial conditions
given at $s=0$:
\begin{equation}
	\xbar\lambda\, \Big\vert_{s=0} = \xbar\epsilon(x^A), \qquad \d_s\xbar
	\lambda\, \Big\vert_{s=0} = \xbar \mu(x^A).
\end{equation}
Both $\xbar \epsilon$ and $\xbar \mu$ are functions on the sphere with a
definite parity: $\xbar\epsilon(-x^A) = \xbar\epsilon(x^A)$ and
$\xbar\mu(-x^A) = -\xbar\mu(x^A)$. The explicit change of parametrisation from
null infinity $\epsilon^{\mathcal J}$ and the Hamiltonian formalism at spatial
infinity $(\xbar \epsilon, \xbar \mu)$ can be obtained using the
expansion in spherical harmonics and the expressions for the functions
$\alpha_l$:
\begin{gather}
	\epsilon^{\mathcal J} = \sum_{l,m} \xbar\lambda_{l,m} 
	Y_{l,m}(x^A), \\
	\xbar\mu = \sum_k \sum_{m=-2k-1}^{2k+1} \xbar\mu_{2k+1,m}\, Y_{2k+1,m}, \quad
	\xbar \epsilon = \sum_k \sum_{m = -2k}^{2k} \xbar\epsilon_{2k,m} \, Y_{2k,m},\\
	\xbar\lambda_{2k+1,m}\, \d_s\alpha_{2k+1}\vert_{s=0} =\xbar\mu_{2k+1,m}, \quad
	\xbar\lambda_{2k,m} \, \alpha_{2k}\vert_{s=0} = \xbar\epsilon_{2k,m}.
\end{gather}
The first few $\alpha_l$ functions can be easily computed
\begin{equation}
	\alpha_0 = 1, \quad \alpha_1 = s + \frac{1-s^2}{2} \log
	\left(\frac{1+s}{1-s}\right), \quad \alpha_2 = 3s^2 -2 +\frac{3s (1-s^2)}{2} \log
	\left(\frac{1+s}{1-s}\right)
\end{equation}
and we can use them to write the first few component of the change of basis:
\begin{equation}
	\xbar\lambda_{0,0} = \xbar\epsilon_{0,0}, \quad \xbar\lambda_{1,m} =
	\frac 1 2 \xbar \mu_{1,m}, \quad \xbar \lambda_{2,m} =  - \frac 1 2 \xbar \epsilon_{2,m}.
\end{equation}

\vspace{5mm}

The action of Lorentz algebra on both parametrizations of the large $U(1)$
gauge transformations can be evaluated using the same strategy.  Lorentz
algebra on the hyperboloid is generated by
\begin{equation}
	\label{eq:hyperbollorentz}
	\mathcal Y^s = - (1-s^2) b, \quad \mathcal Y^A = Y^A  -
	s \xbar D^A b,
\end{equation}
and acts on the $U(1)$ gauge parameter $\xbar \lambda$ through the Lie
derivative. We obtain its action on the gauge parameter at null infinity
$\epsilon^{\mathcal J}$ by taking the limit $s\to 1$:
\begin{equation}
	\delta_{Y,b} \epsilon^{\mathcal J} =\lim_{s \to 1} \delta_{Y,b} 
	\xbar\lambda = (Y^A - \xbar D^A b) \d_A \epsilon^{\mathcal J},
\end{equation}
which is the Lorentz action on the gauge parameter at null infinity
\cite{Barnich:2013sxa}.  The action of Lorentz algebra in the alternative
parametrization can be obtained by evaluation of the transformation laws of
$\xbar\lambda$ and $\d_s \xbar\lambda$ at $s=0$:
\begin{equation}
	\delta_{Y,b} \xbar\epsilon = - b \xbar \mu + Y^A \d_A \xbar
	\epsilon,\quad
	\delta_{Y,b} \xbar \mu = - \xbar D_A(b \xbar D^A\epsilon) + Y^A
	\d_A \xbar \mu.
\end{equation}
This is the action obtained in section \ref{sec:EMhamilSym} which finishes the
proof that the algebra obtained in the hamiltonian formalism is identical to
the one obtained at null infinity.

Note that in the null infinity terminology, the asymptotic symmetry group at spatial infinity is the diagonal subgroup of the groups of angle-dependent transformations at future null infinity and past null infinity.

\end{appendix}

\end{document}